\documentclass[fleqn,usenatbib]{mnras}
\usepackage{newtxtext,newtxmath}

\usepackage[T1]{fontenc}

\DeclareRobustCommand{\VAN}[3]{#2}
\let\VANthebibliography\thebibliography
\def\thebibliography{\DeclareRobustCommand{\VAN}[3]{##3}\VANthebibliography}

\usepackage{graphicx}	
\usepackage[export]{adjustbox}

\usepackage{graphicx}    
\usepackage{dcolumn}     
\usepackage{bm}          
\usepackage{color}
\usepackage{fixltx2e}
\usepackage{tikz}
\usetikzlibrary{positioning}

\usepackage{hyperref}

\usepackage{graphicx}

\usepackage{caption}
\DeclareCaptionFormat{nocaption}{}
\AtBeginDocument{%
}
\usepackage{upquote} 
\usepackage{eurosym} 
\usepackage[mathletters]{ucs} 
\usepackage{fancyvrb} 
\usepackage{grffile} 
\makeatletter 

\usepackage{tikz,xcolor,hyperref}
\definecolor{lime}{HTML}{A6CE39}
\DeclareRobustCommand{\orcidicon}{%
    \begin{tikzpicture}
    \draw[lime, fill=lime] (0,0) 
    circle [radius=0.16] 
    node[white] {{\fontfamily{qag}\selectfont \tiny ID}};    \draw[white, fill=white] (-0.0625,0.095) 
    circle [radius=0.007];    \end{tikzpicture}
    \hspace{-2mm}}
\foreach \x in {A, ..., Z}{%
    \expandafter\xdef\csname orcid\x\endcsname{\noexpand\href{https://orcid.org/\csname orcidauthor\x\endcsname}{\noexpand\orcidicon}}
    }

\title[Eliminating polarization leakage with deep learning]
      {Eliminating polarization leakage effect for neutral hydrogen intensity mapping with deep learning}
\author[Gao, Li, Ni \& Zhang]
  {
  Li-Yang Gao\orcidA{}$^1$,
  Yichao Li\orcidB{}$^1$\thanks{E-mail: liyichao@mail.neu.edu.cn},
  Shulei Ni\orcidC{}$^1$,
   Xin Zhang\orcidD{}$^{1,2,3}$\thanks{E-mail: zhangxin@mail.neu.edu.cn} \\
   $^1$ Key Laboratory of Cosmology and Astrophysics (Liaoning) \& College of Sciences, Northeastern University, Shenyang 110819, China\\
   $^2$ Key Laboratory of Data Analytics and Optimization for Smart Industry (Ministry of Education), Northeastern University, Shenyang 110819, China\\
   $^3$ National Frontiers Science Center for Industrial Intelligence and Systems Optimization, Northeastern University, Shenyang 110819, China
  }

\begin{document}
  \date{\today}
  \label{firstpage}
  \pagerange{\pageref{firstpage}--\pageref{lastpage}}
  \maketitle

  \begin{abstract}
  The neutral hydrogen (HI) intensity mapping (IM) survey is regarded as a promising approach for cosmic large-scale structure studies. 
  A major issue for the HI IM survey is to remove the bright foreground contamination. 
  A key to successfully removing the bright foreground is to well control or eliminate the instrumental
  effects. In this work, we consider the instrumental effects of polarization leakage and use 
  the U-Net approach, a deep learning-based foreground removal technique, 
  to eliminate the polarization leakage effect.
 The thermal noise is assumed to be a subdominant factor compared 
  with the polarization leakage for future HI IM surveys and ignored in this analysis.
  In this method, the principal component analysis (PCA) foreground subtraction is used as a 
  preprocessing step for the U-Net foreground subtraction. 
  Our results show that the additional U-Net processing could either remove the foreground 
  residual after the conservative PCA subtraction or 
  compensate for the signal loss caused by the aggressive PCA preprocessing.
  Finally, we test the robustness of the U-Net foreground subtraction technique and show that 
  it is still reliable in the case of existing constraint error on HI fluctuation amplitude. 
  \end{abstract}


\section{Introduction}
\label{sec1}

Measurements of cosmic large-scale structure (LSS) are
critical for understanding the evolutionary history of the Universe.
In recent decades, the LSS fluctuation of the underlying dark matter
has been explored by mapping the galaxies' distribution in the Universe
using wide-field spectroscopic and photometric surveys
\citep[e.g.][]{Beutler:2011hx,Ross:2014qpa,BOSS:2016wmc}.
The cosmic LSS can also be probed at radio bands by observing
the neutral hydrogen (HI) in galaxies via its 21 cm emission line of 
hyperfine spin-flip transition
\citep[e.g.][]{2004MNRAS.355.1339B,McQuinn:2005hk,Loeb:2008hg}.
Instead of resolving the HI emission line from each distant 
galaxy, the LSS HI survey can be quickly carried out by measuring 
the total HI intensity of many galaxies within a large voxel. 
Such LSS survey technique is known as HI intensity mapping (IM)
\citep{McQuinn:2005hk,Loeb:2008hg,Mao:2008ug,Lidz:2011dx,
Battye:2012tg}, which is ideal for cosmological studies.
\citep[e.g.][]{Xu:2014bya,Bull:2014rha,Yohana:2019ahg,
Zhang:2019dyq,Zhang:2019ipd,Xu:2020uws,Gao:2021xnk,Wu:2021vfz,
Zhang:2021yof,Wu:2022jkf,Jin:2021pcv,Wu:2022dgy,Jin:2020hmc}.

Since the first detection of the cross-correlation function 
between the HI IM survey and the optical galaxy survey using Green Bank Telescope
\citep{Chang:2010jp}, there are a number of measurements of the cross-correlation 
power spectrum between HI IM survey and optical galaxy survey
\citep[e.g.][]{Masui:2012zc,Anderson:2017ert,
2017MNRAS.464.4938W,2021arXiv210204946W,2022arXiv220201242C}. 
Moreover, several radio telescopes and interferometers specially designed 
for HI IM survey are either under construction or collecting data, i.e.,
the Tianlai project \citep{2012IJMPS..12..256C,2020SCPMA..6329862L,2021MNRAS.506.3455W,2022MNRAS.517.4637P,2022RAA....22f5020S}, 
the Canadian Hydrogen Intensity Mapping Experiment \citep[CHIME,][]{2014SPIE.9145E..22B}, 
the Baryonic Acoustic Oscillations from Integrated Neutral Gas Observations \citep[BINGO,][]{2013MNRAS.434.1239B} and the
Hydrogen Intensity and Real-Time Analysis experiment ~\citep[HIRAX,][]{2016SPIE.9906E..5XN}.
The HI IM survey is also proposed as a major cosmological probe using 
the Square Kilometre Array \citep[SKA,][]{2015aska.confE..19S}, 
and its pathfinder array, MeerKAT \citep{2017arXiv170906099S,
2021MNRAS.501.4344L,2021MNRAS.505.3698W}.
Recently, \citet{2022arXiv220601579C} reported the detection of cross-correlation power spectrum
of the MeerKAT HI IM survey and the WiggleZ optical galaxy survey.
Meanwhile, using the MeerKAT interferometric observations, \citet{2023arXiv230111943P} reported the HI IM auto-correlation power spectrum detection on Mpc scales. 

However, there still are challenges for HI IM cosmological studies.
The auto-correlation power spectrum of the HI fluctuation on large scales has not yet been detected
with the IM survey at this time \citep{2013MNRAS.434L..46S}.
The main challenge for this observation method is to extract the HI LSS signal 
from the bright foreground contamination, e.g. the Galactic synchrotron emission, the Galactic free-free emission,
and the extragalactic radio sources \citep{2021MNRAS.504..208C,Spinelli:2021emp,Wolz:2015sqa}.
Although this task may be impossible at first, it has been proposed 
that the smooth frequency dependence of the foreground can be used to separate them from
the faint HI LSS signal \citep{2009ApJ...695..183B,Ansari:2011bv}.
For example, \citet{Ansari:2011bv,Bigot-Sazy:2015jaa} used non-blind parametric fitting to remove the foregrounds.
\citet{Shaw:2013wza} implemented the Karhunen Lo{\`e}ve (KL) transform for 
foreground subtraction, which extracts the HI signal by modeling the statistical 
properties of various components of the sky (e.g., signal, foreground, and noise).
However, the instrumental effects, existing in all radio telescope systems, significantly increase 
the difficulty and complexity of such components separation technique \citep{Matshawule:2020fjz,2022MNRAS.509.2048S}.
To address the instrumental effects, a couple of blind foreground subtraction approaches have been developed, 
such as the Principal Component Analysis \citep[PCA,][]{Masui:2012zc}, the Fast Independent Component Analysis
\citep[\textsc{FastICA},][]{Chapman:2012yj,Cunnington:2019lvb,Wolz:2013wna,761722}, and Generalised Morphological Component Analysis \citep[GMCA,][]{4337755}. Especially, the PCA and FastICA methods are already applied to the 
real HI IM survey data \citep{Masui:2012zc,Anderson:2017ert,2021arXiv210204946W,Patil:2017zqk}.
We emphasize that our list of methods is not an exhaustive list and there are probably many more methods that can be used for foreground removal of HI IM.

The instrumental effect complicates the structure of the foreground spectrum.
Thus, eliminating the instrumental effect is crucial for efficiently removing the foreground.
When considering the instrumental effect, the rapidly growing deep learning
algorithm can play a critical role in foreground subtraction \citep{Goodfellow-et-al-2016}.
Deep learning has been widely used in cosmology and radio astronomy, such as using deep learning to predict 
the cosmological structure formation \citep{He:2018ggn}, generating HI mock signals \citep{Wadekar:2020ear}, 
simulating and exploring cosmic dawn and epoch of reionization
\citep{Gillet:2018fgb,Kwon:2020ual,List:2020qdz,Mangena:2020jdo,Villanueva-Domingo:2020wpt}, etc.
In the meanwhile, deep learning can also be used in foreground subtraction \citep{Li:2019znt}.
\citet{Makinen:2020gvh} reported a foreground subtraction method based on a convolutional neural network (CNN) 
architecture based on a $3$D U-Net model \citep{Kohl2018APU}.
\citet{Ni:2022kxn}, for the first time, proposed a deep learning approach to eliminate the primary 
beam effect in HI IM foreground subtraction.

In addition to the primary beam effect, the instrumental polarization leakage 
also needs to be properly controlled \citep{Nunhokee:2017aaj,Bhatnagar:2001mh}.
Although the polarization of the HI emission is negligible, the foreground contamination, 
especially the bright synchrotron emission is polarized. Due to the imperfect 
calibration, the polarized foreground emission can leak into the total intensity.
Such leakage could induce extra power and complicate the foreground spectrum. 
\citet{Liao:2016jto} provided a detailed polarization calibration for GBT HI IM survey
and showed that the polarization calibration error on boresight is well controlled.
However, the polarization leakage can still be identified via the polarized beam pattern measurements.
A few more studies are focusing on simulating polarization leakage for different HI experiments 
\citep{Shaw:2014khi,https://doi.org/10.1002/asna.200610566,Schnitzeler:2008ga,Wolleben:2005td,
Nunhokee:2017aaj}.
Eliminating the polarization leakage from the full beam pattern is
important for separating the foreground from the HI signal.

In this paper, we further extend the deep-learning foreground subtraction approach
to eliminate the instrumental effect of polarization leakage.
This paper is organized as follows. 
In Section \ref{sec2}, we present the relevant formulae and the simulation of the sky maps.
Section \ref{sec3} describes the methods used to extract the HI signal. 
The results are discussed accordingly in Section \ref{sec4} and
the conclusions are presented in Section \ref{sec5}.

\section{Simulation}
\label{sec2}

The deep learning network is trained using a set of simulated sky maps,
which are generated using the open-source package, 
{\tt CRIME}~(\url{http://intensitymapping.physics.ox.ac.uk/CRIME.html})
 \citep{Alonso:2014sna}. 
The sky maps include three main components, 
i.e. the HI brightness contrast, the foreground, and the leakage from the polarized foreground. 
In this work, we assume that the thermal noise levels of future HI experiments are subdominant 
with respect to the foreground residual. The noise level may also be lower than the HI fluctuation 
on large scales with sufficient long integration time for future large radio telescopes or interferometer arrays \citep{Li:2017jnt}.
As shown in the literature \citep[e.g.,][]{Makinen:2020gvh}, as long as samples with this noise level are included in the training set, the U-Net can handle different noise realizations at the same noise level.
To simplify the simulation analysis and highlight the effect of the polarization leakage, 
the thermal noise is ignored in this work.
A brief overview of the simulation processing is provided below.

\subsection{HI brightness contrast}
\label{sec:simhi}

The HI brightness temperature contrast at redshift $z$ is expressed as
\begin{equation}
\begin{aligned}
T_{\rm HI}(z,\hat{\mathbf{n}}) 
& = 190.55~{\rm mK}\frac{\Omega_{\rm b} h (1+z)^2 
x_{\rm HI}(z)}{\sqrt{\Omega_{\rm m} (1+z)^3 + \Omega_\Lambda}} 
(1+\delta_{\rm HI}(z,\hat{\mathbf{n}})) \\
&= \bar{T}_{\rm HI}(z)\left( 1 +  \delta_{\rm HI}(z,\hat{\mathbf{n}})\right),
\label{T21}
\end{aligned}
\end{equation}
where 
$h \equiv H_0/(100~{\rm km~s}^{-1}~{\rm Mpc}^{-1})$ is the dimensionless Hubble constant,
$x_{\rm HI}(z)$ is the neutral hydrogen mass fraction in terms of the total baryons,
$\Omega_{\rm b}$, $\Omega_{\rm m}$ and $\Omega_\Lambda$ denote the present baryon, 
total matter and dark energy density fractions, respectively, 
and $\delta_{\rm HI}(z, \hat{\mathbf{n}})$ represent the redshift-space HI density contrast.

The full sky maps of the HI brightness contrast are simulated using a set of 
lognormal realizations with the tomographic angular power spectrum
\begin{equation}
C_\ell^{\rm HI}(z_1, z_2) = \frac{2}{\pi} \int {\rm d}k\,k^2 
P_{\rm DM}(k) W_{\ell, z_1}(k) W_{\ell, z_2}(k),
\end{equation}
where $P_{\rm DM}(k)$ is the present underlying dark matter power spectrum
and $W_{\ell, z_i}(k)$ denotes the window function of a $z$-shell 
with the mean redshift at $z_i$,
\begin{equation}
W_{\ell, z_i}(k) = \int {\rm d}z \bar{T}_{\rm HI}(z) \phi(z_i, z) D(z) 
\left(b(z)j_\ell(k\chi) - f(z) j''_\ell(k\chi) \right),
\end{equation}
where $j_\ell$ is the $\ell$-th spherical Bessel function, 
$D(z)$ is the growth factor, $f(z)$ is the growth rate, $b(z)$ 
is the linear bias of the HI density contrast with respect to the
dark matter density field, and 
$\phi(z_i, z)$ is the redshift selection function centering at $z_i$.
In this work, the simulated maps are generated across the uniform frequency
slices according to $z = \nu_0/\nu - 1$, where $\nu_0 = 1420.406\,{\rm MHz}$ 
is the HI emission rest frame frequency. 
The cosmological parameters are fixed to the best-fit values from \cite{Planck:2018vyg}.

\subsection{Unpolarized foreground components}
\label{sec:Unpolar}

We first consider the total intensity contamination from the foreground emission.
The total foreground intensity maps are simulated using two different methods
according to their different angular distribution properties, i.e. anisotropic
and isotropic distribution. 

The Galactic synchrotron emission has a highly anisotropic angular structure.
In the simulations, the anisotropic Galactic synchrotron emission is
simulated by interpolating the Haslam map's brightness temperature at $408$ MHz \citep{Haslam:1982zz} to the required frequencies via
\begin{equation}
T_{\rm syn}(\nu, \hat{\mathbf{n}}) = T_{\rm Haslam}(\hat{\mathbf{n}}) 
\left(\frac{408\,{\rm MHz}}{\nu}\right)^{\gamma(\hat{\mathbf{n}})}
\end{equation}
where $\gamma(\hat{\mathbf{n}})$ is the direction-dependent spectra index. 
The $\gamma(\hat{\mathbf{n}})$ values are read from the Planck Sky Model \citep[PSM][]{Delabrouille:2012ye}.

In addition to the Galactic synchrotron emission, a set of isotropic foreground
emissions are generated following the method proposed by Santos, Cooray, and Knox (SCK)
\citep{Santos:2004ju}. In this case, the foreground maps are simulated with
the Gaussian random realization based on the angular power spectrum
\begin{equation}
C_\ell(\nu_1,\nu_2) = A \left(\frac{\ell_{\rm ref}}{\ell}\right)^\beta \left(\frac{\nu^2_{\rm ref}}{\nu_1 \nu_2}\right)^\alpha \rm exp \left(-\frac{\rm log^2(\nu_1/\nu_2)}{2\xi^2}\right)
\label{C_l},
\end{equation}
where $\xi$ is the frequency-space correlation length of the emission.
The extragalactic point sources include both the nearby
and the high-redshift radio galaxies. The angular distributions of both are
assumed to be isotropic. 
The Galactic free-free emission is related to the electron distribution and 
not homogeneous. However, the free-free emission
is a subdominant component across the simulation frequency range. 
The Galactic free-free, together with the extragalactic free-free emission, can be
approximated as the isotropic foreground.
Besides, the Galactic synchrotron emission at small scales below the angular resolution 
of the Haslam map is also simulated as an isotropic component. 
The parameters used in the SCK method for each of the isotropic components 
are listed in Table \ref{tab_cl}.

\begin{table}
\begin{center}
\renewcommand{\arraystretch}{1.5}
\caption{\label{tab_cl}  Parameters used in foreground $C_\ell(\nu_1,\nu_2)$ model adapted from \citet{Santos:2004ju} and \citet{Alonso:2014sna} for the pivot values $\rm \ell_{ref} = 1000$ and $\rm \nu_{ref} = 130~MHz$.}
\begin{tabular}{c c c c c}
\hline
Foreground & $\rm A~(mK^2)$ & $\beta$ & $\alpha$ & $\xi$ \\
\hline
Galactic synchrotron & $700$ & $2.4$ & $2.80$ & $4.0$\\
Point sources & $57$ & $1.1$ & $2.07$ & $1.0$\\
Galactic free-free & $0.088$ & $3.0$ & $2.15$ & $35$\\
Extragalactic free-free & $0.014$ & $1.0$ & $2.10$ & $35$\\
\hline
\end{tabular}
\end{center}
\end{table}

\subsection{Leakage from polarized foreground}
\label{sec:Leakage}


\begin{figure*}
\centering
\includegraphics[width=0.95\textwidth]{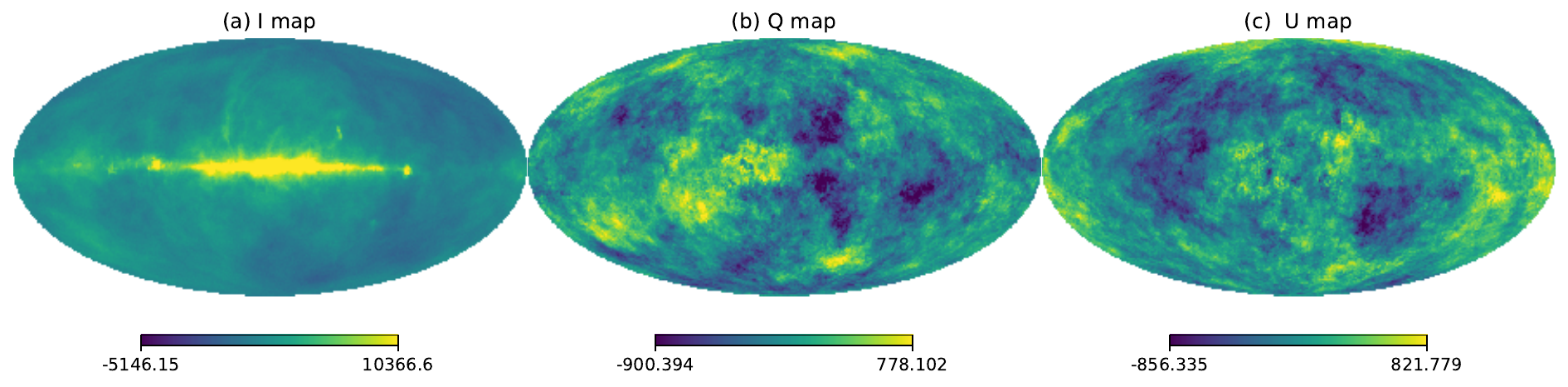}
\caption{\label{polar_map}
The sky maps of simulated Galactic synchrotron emission at the frequency of 980 MHz. 
The stocks I, Q, and U components are shown in the (a), (b), and (c) panels, respectively. 
All the maps are shown in units of mK.
All of the maps in this work are drawn in Galactic coordinates using the Mollweide projection.
} \end{figure*}

For a perfectly calibrated telescope, the measured intensity is only 
sensitive to the total intensity, rather than the different polarization modes Q or U. 
However, for more general cases of a non-ideal telescope, there is always leakage 
between different polarization modes. Such polarization leakage can be described 
via a $4\times4$ polarization calibration error matrix
\citep{2004ApJS..152..129V,Liao:2016jto,Nunhokee:2017aaj}.
The polarization calibration error matrix needs to be measured via polarization calibration observation.
To simplify the analysis, we only consider the intensity leakage from the polarized 
foreground components, i.e. the Stokes Q and U maps of the foreground 
(the Stokes V map is ignored as the circular polarization is assumed to be negligible).
We also assume a constant polarization leakage fraction across the frequency band.
Thus the polarization leakage is expressed as
\begin{equation}
T_{\rm leak}=\epsilon_Q T^Q_{\rm FG} + \epsilon_U T^U_{\rm FG},
\label{T_leak}
\end{equation}
where $T^Q_{\rm FG}$ and $T^U_{\rm FG}$ denote the Stocks Q and U maps of the polarized 
foreground, respectively, while $\epsilon_Q$ and $\epsilon_U$ represent the constant intensity 
leakage fractions from $T^Q_{\rm FG}$ and $T^U_{\rm FG}$. 
Because synchrotron radiation is substantially stronger than other foreground components,
we ignored the polarized components from other foreground emissions in our simulations.

The Galactic synchrotron emission is known to be partially linearly polarized
\citep{1970ranp.book.....P,1986rpa..book.....R} 
and its polarization angle rotated due to the frequency-dependent 
Faraday rotation effect \citep{Brentjens:2005zc,2010MNRAS.409.1647J,2013ApJ...769..154M}.
Such effect, combined with the polarization leakage of the non-ideal telescope, significantly 
complicates the foreground frequency structure and makes foreground subtraction difficult
\citep{2021MNRAS.504..208C}.
We use {\tt CRIME} package to generate the Galactic synchrotron Stocks Q and U maps 
at each frequency, which is constructed in the Faraday rotation measure space.
The details for determining the polarized synchrotron emission 
using Faraday depth measurements are given in
\citet{Alonso:2014sna} and \citet  {Oppermann:2014cua}.
In Fig.~\ref{polar_map}, the simulated synchrotron stocks I, Q, and U maps are shown in the left, middle, and right panels, respectively. 
 
Using the simulated Q and U maps, we vary the telescope polarization leakage fraction.
We know from \citet{deVilliers:2022vhg} that the leakage from the I map to both the Q map and the U map is roughly less than $2\%$.
The Mueller matrix is composed of multiple Ermey matrices multiplied by each other \citep{Liao:2016jto,Britton:1999mma}, so we further consider the simple case that the Mueller matrix is a symmetric matrix.
On this premise, we give the intensity leakage fractions.
Firstly, we assume no polarization leakage and set both $\epsilon_U=0$ and $\epsilon_Q=0$.
Then we keep $\epsilon_U=0$ and set $\epsilon_Q=\{0.5\%, 1.0\%, 2.0\%\}$. Finally, we 
set $\epsilon_U=1.0\%$ and $\epsilon_Q=\{0.0\%, 0.5\%, 1.0\%, 2.0\%\}$.

\subsection{Combination of the sky maps}
\label{sec:plist}

The simulated sky map is the combination of different components, i.e. 
the HI signal, the foreground components, and the leakage of the polarized synchrotron
emission,
\begin{equation}
T_{\rm sky}(\nu, \hat{\mathbf{n}})=T_{\rm HI}+T_{\rm unpol}+T_{\rm leak}.
\label{T_sim}
\end{equation}


In this work, we focus on the HI survey of the late-time Universe.
Thus, the sky maps are simulated in the frequency range
of $\rm 970~MHz$ to $\rm 1023~MHz$, corresponding to the MeerKAT L-band HI IM survey
\citep{2017arXiv170906099S}.
Each map contains $64$ frequency channels.
The maps are also simulated using different polarization leakage settings.
The different polarization leakage settings are list as below,
\begin{itemize}
\item $(\epsilon_Q, \epsilon_U) = ( 0.0\%, 0.0\%)$;
\item $(\epsilon_Q, \epsilon_U) = ( 0.5\%, 0.0\%)$;
\item $(\epsilon_Q, \epsilon_U) = ( 1.0\%, 0.0\%)$;
\item $(\epsilon_Q, \epsilon_U) = ( 2.0\%, 0.0\%)$;
\item $(\epsilon_Q, \epsilon_U) = ( 0.0\%, 1.0\%)$;
\item $(\epsilon_Q, \epsilon_U) = ( 0.5\%, 1.0\%)$;
\item $(\epsilon_Q, \epsilon_U) = ( 1.0\%, 1.0\%)$;
\item $(\epsilon_Q, \epsilon_U) = ( 2.0\%, 1.0\%)$.
\end{itemize}


\section{Method}
\label{sec3}


\subsection{Traditional blind foreground subtraction method}
\label{sec3.1}

There are different traditional foreground subtraction methods, e.g.
PCA, fastICA, and GNILC, that have been tested with various HI IM surveys 
\citep{Masui:2012zc,Anderson:2017ert,
2017MNRAS.464.4938W,2021arXiv210204946W,2022arXiv220201242C,2022arXiv220601579C,Olivari:2015tka}.
Such blind foreground subtraction approaches can eliminate complex foreground 
components without requiring a thorough understanding of the systematic effect.
Simulations also show that the blind foreground subtraction approaches produce
similar results \citep{2021MNRAS.504..208C}.
In this work, we investigate the performance of the PCA approach in the presence
of various polarization leakages.

The PCA approach is accomplished by doing the singular value decomposition (SVD)
of the frequency-frequency covariance matrix of each line-of-sight spectra.
Because the foreground components are highly correlated across frequencies
as compared to the HI fluctuation, the foreground contamination is represented 
by the first few SVD modes with bigger singular values.
The number of SVD modes to be removed depends on the complexity of the foreground spectra.
A couple of effects, e.g. the antenna primary beam sidelobe
\citep{Matshawule:2020fjz}, satellite communication \citep{2018MNRAS.479.2024H}, 
polarization leakage \citep{2021MNRAS.504..208C}, can significantly complicate
the foreground spectra. 
As a result, removing a greater number of modes is required to clear up the 
contamination. However, such aggressive foreground subtraction always
results in significant signal loss \citep{Switzer:2015ria}. 

Instead of applying aggressive foreground subtraction, we propose a 
light PCA foreground subtraction with additional subtraction based on the
deep learning method.
By subtracting only a few modes, the primary foreground contamination can be
removed leaving the residual dominated by the systematic effect.
\citet{Ni:2022kxn} explored such foreground subtraction approaches  
with a deep learning architecture to eliminate the primary beam effect in 
foreground subtraction.
The PCA subtraction, on the other hand, is known as a crucial 
preprocessing step for the deep learning approach \citep{Makinen:2020gvh}.

\subsection{The U-Net architecture}
\label{sec3.2}

\begin{figure*}
\centering
\includegraphics[width=0.90\textwidth]{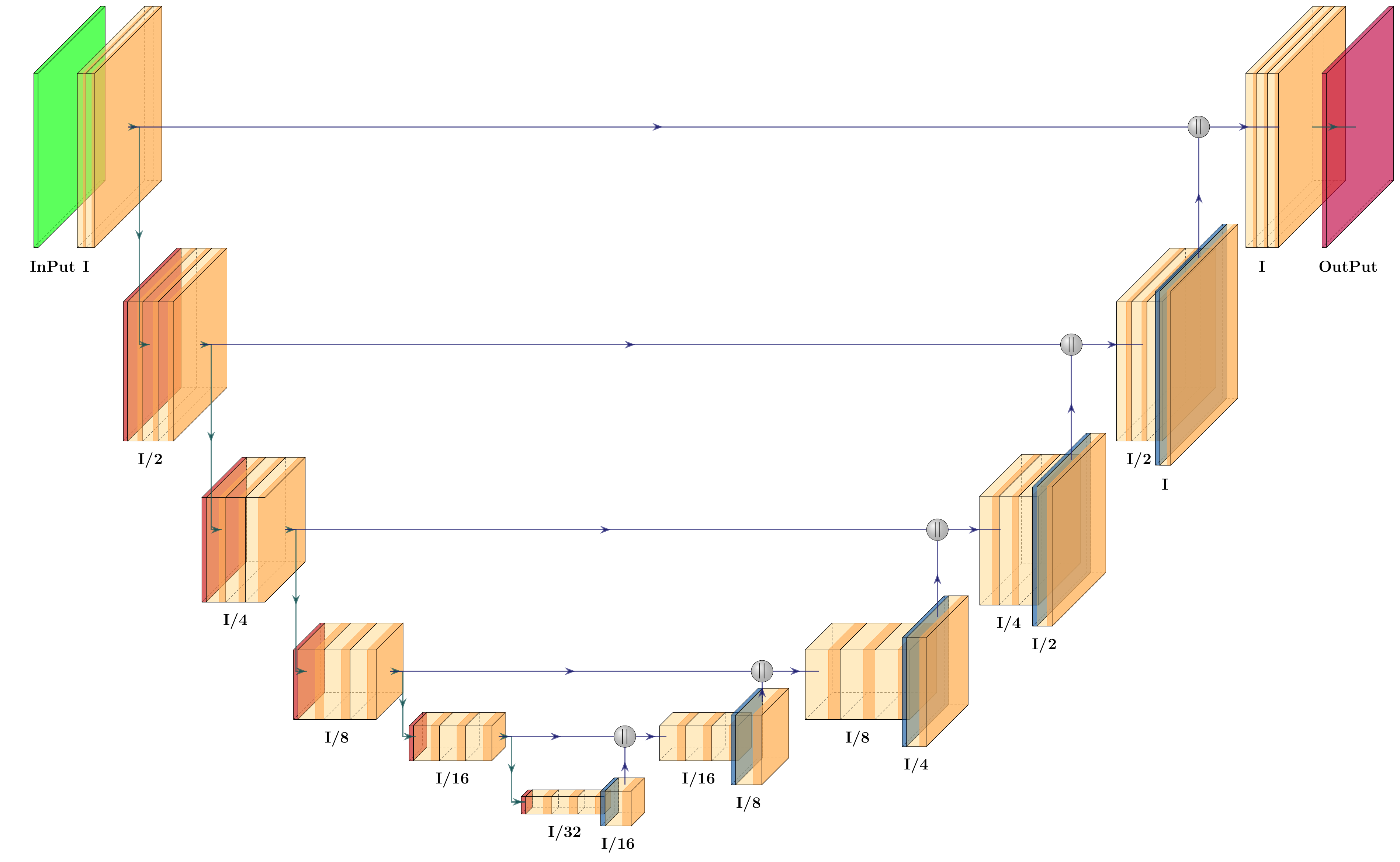}
\caption{\label{unet}The training process of CNN with U-Net architecture. Each color represents a structure in the U-Net network, where yellow cubes represent the convolutional layers and ReLU sections, red cubes represent pooling layers in down-sampling, blue cubes represent the transposed convolutional layers, and grey spheres represent connection layers. The green and purple squares at the beginning and end of this figure represent the input and output, respectively.}
\end{figure*}

The $5$-layers processing of the U-Net architecture is illustrated in 
Fig.~\ref{unet}.
The square on the left side of the image is the cube of our input PCA-processed sky maps, while the square on the right side of the image is the data from our deep learning output.
The U-Net architecture consists of down-sampling processing, 
shown in the left-hand-side of Fig.~\ref{unet}, and up-sampling processing,
in the right-hand-side of Fig.~\ref{unet}, respectively.  
The down-sampling processing is a common convolutional network composed
of repeated convolutional applications, each followed by a rectified linear
unit (ReLU) and a maximum set operation.
The convolutional layer and ReLU are illustrated with yellow boxes 
and the maximum set operation is shown with red boxes.
With such down-sampling processing, the structure information of the input image
are contracted into a few features. 
An important improvement over the common convolution network is the 
supplementation of the up-sampling processing.
The blue part represents the transposed convolution in the
up-sampling process, while the grey sphere is the connection layer.
The up-sampling processing combines features and spatial information
through a sequence of transposed convolutions, as well as the concatenation
with the down-sampled high-resolution outputs.
With such up-sampling processing, the network can learn to assemble a 
precise HI fluctuation across the sky.
It is worth noting that for the whole learning process, 
we use the {\tt AdamW} \citep{Loshchilov2017FixingWD} optimizer to obtain
a reasonable stepwise decreasing learning rate.

In this work, we use the U-Net architecture developed by \citet{DBLP:journals/corr/RonnebergerFB15}.
Currently, the U-Net package can only deal with the image data in a cube.
Our full sky simulation maps need to be split into a couple of 
small patches.

\subsubsection{Dataset assembly}
\label{Dataset}
The simulated full sky maps are pixelized using the {\tt HEALPix} pixelization
scheme with $N_{\rm side}=256$. 
As discussed in Section~\ref{sec3.1}, a few of the SVD modes are removed at the beginning
to reduce the dynamic range of the simulated maps.
We follow the same partition strategy as used in \citet{Ni:2022kxn}.
The SVD-modes-removed maps are split into $192$ small patches across the full
frequency channels, each with $64\times64$ pixels and $64$ frequencies.
The $64\times64\times64$ cubes are used as the input cubes for the U-Net network.

For each polarization leakage setting, we generate $5$ different mock maps. 
$3$ of the $5$ mock maps, a total of $576$ cubes, are used as the training 
sets, and the rest $2$ mock maps are used as the verification and testing 
sets, respectively.

Finally, the foreground cleaned sky patches are reconstructed into the full sky maps
using the same {\tt HEALPix} pixelization scheme as the input maps.
We still follow the construction method implemented in \citet{Ni:2022kxn}.


\subsubsection{Loss function}
\label{Loss_function}
The selection of the loss function is crucial for efficient network training.
Since the standard Mean Square Error (MSE) loss function is known to be unstable
in dealing with image problems \citep{Makinen:2020gvh}, and the Log-Cosh loss function is smoother than MSE.
Besides, the Log-Cosh loss function is not easily influenced by anomalous points.
Therefore, we use the Log-Cosh loss function in this work:
\begin{equation}
\mathcal{L}=\sum\limits_{i} \log\cosh(p_i - t_i),
\label{loss_fun}
\end{equation}
where $p_i$ and $t_i$ represent the predicted outcome and the 
true signal of the $i$-th voxel, respectively.

\subsubsection{Hyperparameter selection}
\label{Hyperparameter_selection}
\begin{table}
\begin{center}
\caption{\label{Hyper_unet}Description of the hyperparameters in the U-Net architecture design.}
	\begin{tabular}{llcc}
		\hline
		Hyperparameter                                             & Optimum value\\
		\hline
		$n_{\rm block}$~(number of convolutions for each block)    & 3           \\
		$n_{\rm down}$~(number of down-convolutions)               & 5           \\
		$n_{\rm tc}$~(number of transpose convolutions)            & 4           \\
		batchnorm~(batch normalization for given layer)            & Ture*       \\
		$\omega$~(weight decay for optimizer)                      &$10^{-5}$    \\
		batch size~(number of samples per gradient descent step)   & 16          \\
		$n_{\rm filter}$~(initial number of convolution filters)  & 36          \\
		$\Omega$~(optimizer for training)                          & AdamW       \\
		$\eta$~(learning rate for optimizer)                       & $10^{-3}$   \\
		$\beta_{\rm mom}$~(batch normalization momentum)           & $0.02$      \\
		\hline
	\end{tabular}
\end{center}
\end{table}

\begin{figure}
\centering
\includegraphics[width=0.45\textwidth]{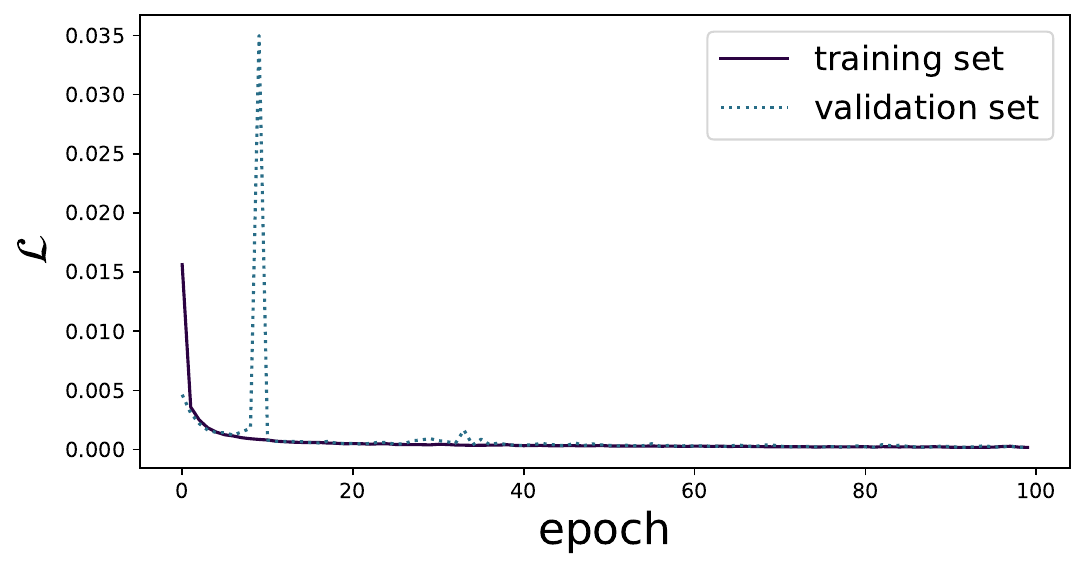}
\caption{\label{loss}
LogCosh loss evolves with the number of epochs. The solid line represents the evolution of the loss in the training set and the dashed line represents the evolution of the loss in the validation set.}
\end{figure}

To optimize the network, the hyperparameters are fine-tuned.
In order to speed up the computation of the neural network,
we increase the learning rate $\eta$ to $10^{-3}$ 
and increase weight decay for optimizer $\omega$ to $10^{-5}$.
In Fig. \ref{loss}, we tested the effect of different epoch numbers. We find that the loss function does not change significantly after the epoch number reaches $20$. 
So in this work, we used $\rm epoch = 20$ for efficiency.
According to the default setting, the kernel size is set to $3$.
We perform batch normalization for the given layer and batch normalization momentum ${(\beta_{\rm mom}) = 1}$.
Besides, we use ${\rm stride}=1$ in the convolutions and ${\rm stride}=2$
in the transpose convolution, respectively.
Due to the limitation of computer arithmetic, we set the batch size to $16$.
Since we want the neural network to be able to learn more structural information about the foreground, there is no doubt that we can improve the number of down-convolutions ($n_{\rm down}$) and the initial number of convolution filters ($n_{\rm filter}$).
Due to the limitation of GPU memory and the little influence of $n_{\rm down}$ and $n_{\rm filters}$ on the results, we set them as $5$ and $36$, respectively.
Therefore, because of the limitations of the neural network structure, the number of transpose convolutions ($n_{\rm tc}$) equals $4$.
Relevant hyperparameters are represented in Table \ref{Hyper_unet}.

\section{Results and Discussion}
\label{sec4}

\subsection{Results with only the PCA approach}
\label{R1}

\begin{figure*}
\centering
\begin{minipage}[t]{0.48\textwidth}
\centering
\includegraphics[width=\textwidth]{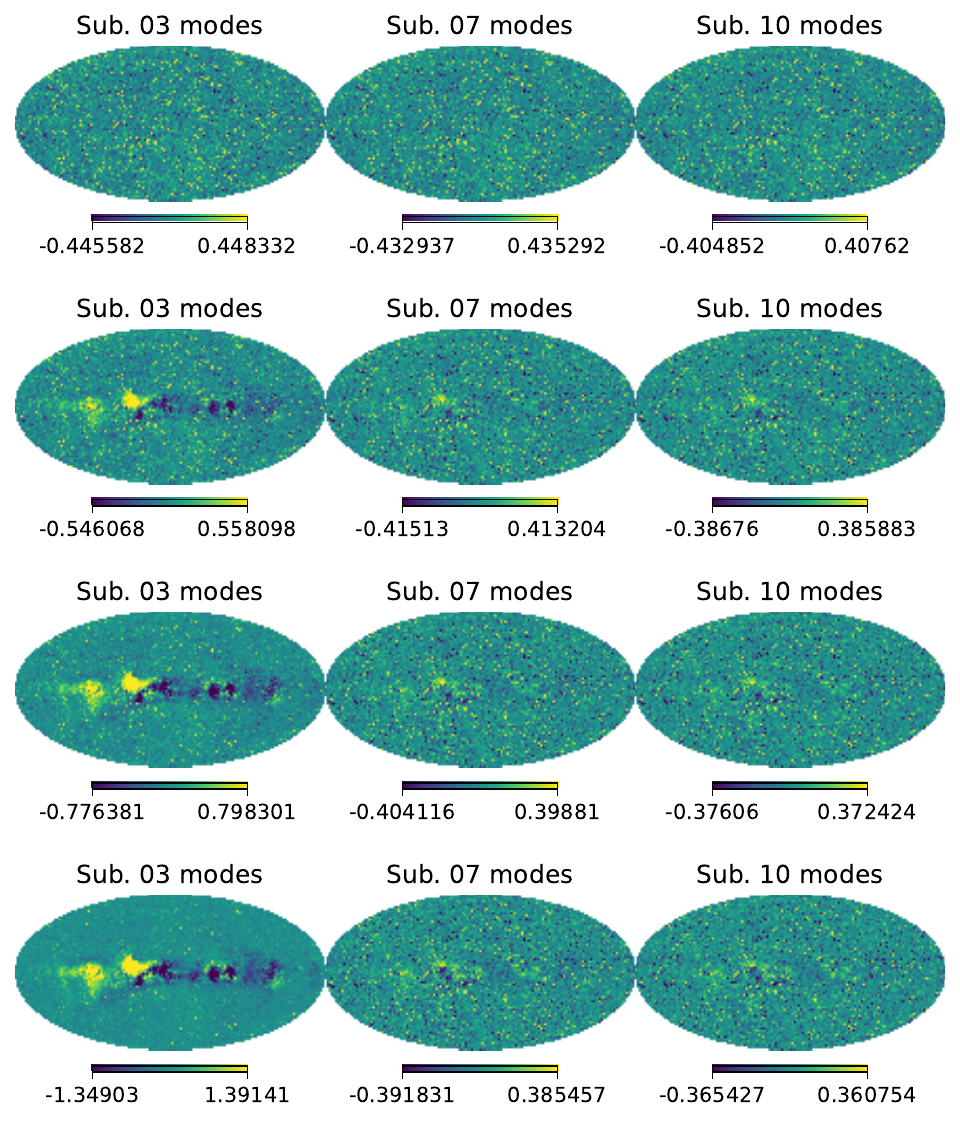}
\caption{
The PCA-only cleaned sky maps.
From top to bottom, the maps with polarization leakage fraction of 
$(\epsilon_Q, \epsilon_U) = ( 0.0\%, 0.0\%)$,
$( 0.5\%, 0.0\%)$, $( 1.0\%, 0.0\%)$ and $( 2.0\%, 0.0\%)$
are shown in different rows.
The maps with $3$, $7$ and $10$ SVD modes subtracted are shown in the 
left, middle and right panels, respectively.
All the maps are in units of mK.}\label{fig:mappca}
\end{minipage} \hfill
\begin{minipage}[t]{0.48\textwidth}
\centering
\includegraphics[width=\textwidth]{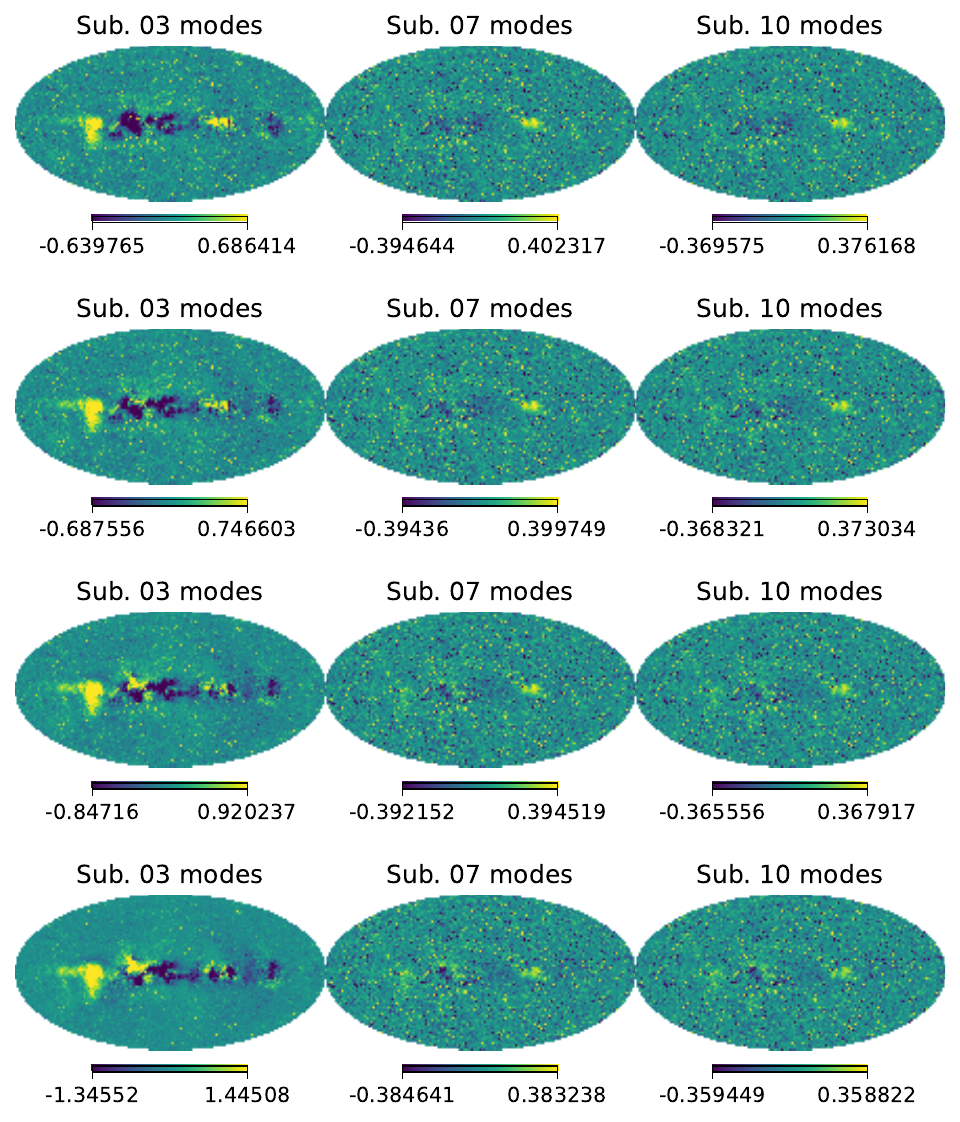}
\caption{Same as Fig.~\ref{fig:mappca} but with polarization leakage fraction
of $(\epsilon_Q, \epsilon_U) = ( 0.0\%, 1.0\%)$,
$( 0.5\%, 1.0\%)$, $( 1.0\%, 1.0\%)$ and $( 2.0\%, 1.0\%)$
are shown in different rows.
All the maps are in units of mK.}\label{fig:mappcaU}
\end{minipage}
\end{figure*}

\begin{figure*}
\centering
\includegraphics[width=0.9\textwidth]{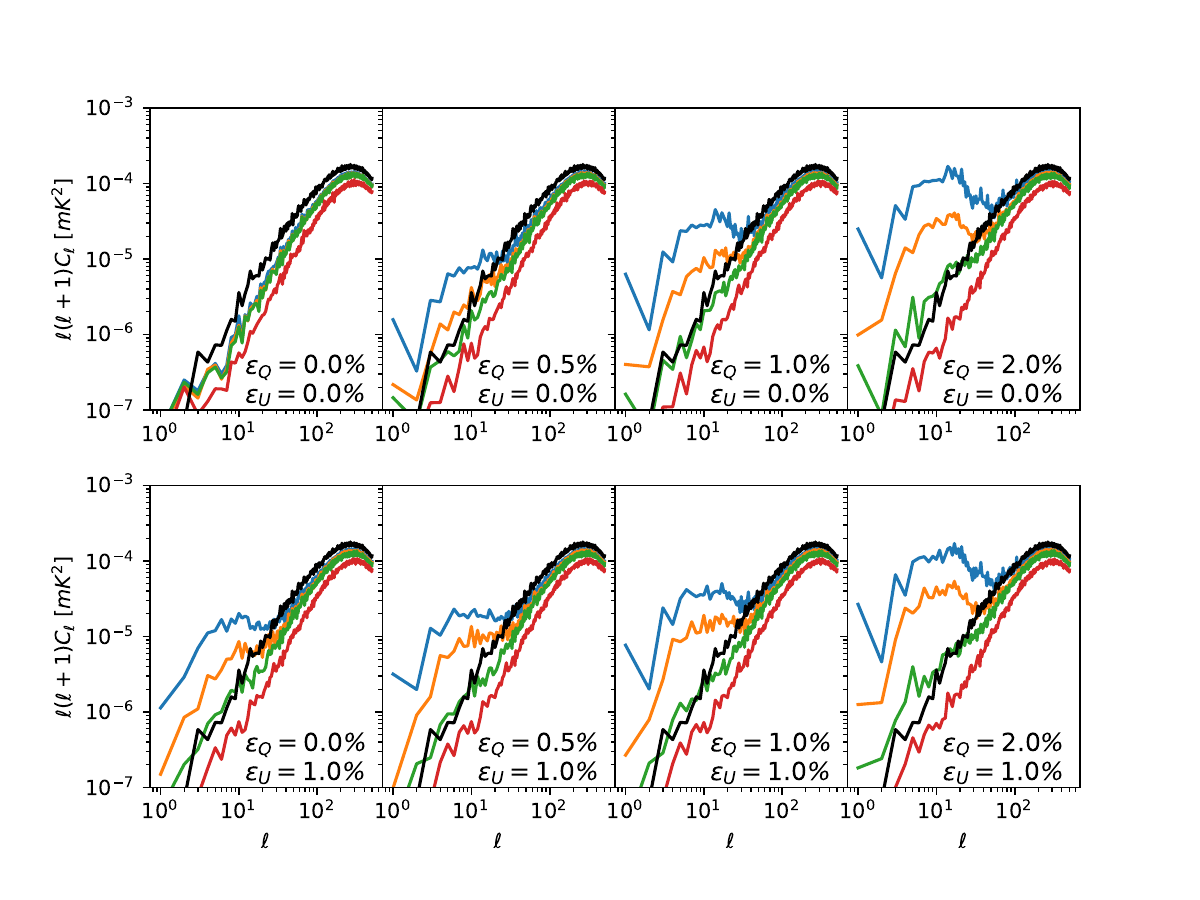}
\caption{\label{PCAcl}
Angular power spectra of PCA-only cleaned sky maps. 
The results with different polarization leakage parameters are shown in different panels.
Columns 1 to 4 represent $\epsilon_Q$ taken as $0.0\%$, $0.5\%$, $1.0\%$, and $2.0\%$, respectively. 
The upper and lower panels represent $\epsilon_U$ taken as $0.0\%$ and $1.0\%$, respectively.
The black lines represent the sky maps of the input HI signal.
The blue, yellow, green, and red lines represent the power spectrum after subtracting the first 3, 4, 5, and 10 SVD modes,
respectively.}
\end{figure*}

\begin{figure}
\centering
\includegraphics[width=0.48\textwidth]{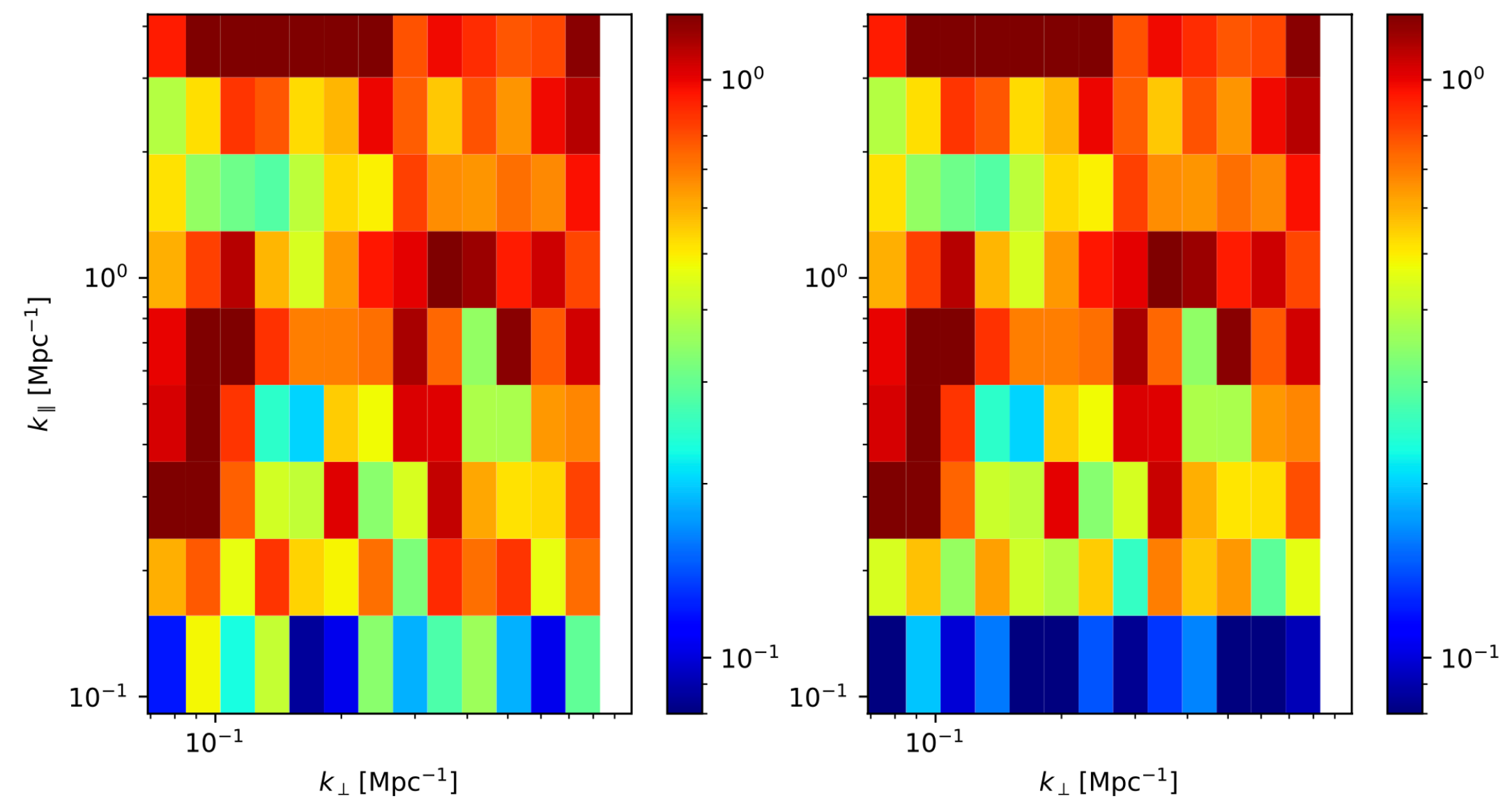}
\caption{\label{2d}
The ratio of the $2$D power spectrum with respect to the input HI signal. 
Assuming no polarization leakage, i.e. $(\epsilon_Q, \epsilon_U) = ( 0.0\%, 0.0\%)$, 
the results of PCA-only foreground cleaned maps with $3$ and $4$ modes subtraction 
are shown in the left and right panels, respectively.}
\end{figure}

We first test the foreground subtraction with only the PCA approach. 
The foreground cleaned sky maps at $980$ MHz with only considering the polarization leakage from the Q map 
are shown in Fig.~\ref{fig:mappca} and with additional leakage from the U map are 
shown in Fig.~\ref{fig:mappcaU}.
In Fig.~\ref{fig:mappca} and Fig.~\ref{fig:mappcaU}, the results with different
polarization leakage levels are shown in different rows. In each row, the maps with
$3$, $7$, and $10$ SVD modes subtracted are shown in the left, middle, and right panels. 
All the maps are shown in Galactic coordinates and in units of mK. 
The amplitude dynamic range of the map is truncated 
between $\pm 3 \sigma_{\rm map}$, where $\sigma_{\rm map}$ is the rms of the whole map. 

The top row of Fig.~\ref{fig:mappca} shows the cleaned sky map without any
polarization leakage. Clearly, without polarization leakage, 
there is no obvious foreground residual structure after subtracting $3$ SVD modes.
We also find that even without any polarization leakage, the foreground subtraction
requires at least $3$ SVD modes. This is consistent with the 
foreground simulation analysis in the literature \citep[e.g.,][]{2022MNRAS.509.2048S}.
When considering the polarization leakage, we assume that any foreground residuals existing
in the SVD modes beyond the first 3 are due to the leakage. 
Such foreground residuals are further eliminated with an aggressive SVD mode subtraction
in most of the HI IM experiments data analysis.
In this work, we subtract the first 3 SVD modes in the pre-processing, 
and adopt the U-Net to eliminate the polarization leakage-induced foreground residuals.

The maps in the second, third, and fourth rows of Fig.~\ref{fig:mappca} represent 
the results with $0.5\%$, $1.0\%$, and $2.0\%$ polarization leakage from the Q map.
The maps in Fig.~\ref{fig:mappcaU} show the results with the same 
Q map polarization leakage level as the same row in Fig.~\ref{fig:mappca}, 
but with additional $1.0\%$ polarization leakage from U map.
With the polarization leakage injection, the maps with only 3 SVD modes subtracted 
show significant foreground residual structure. Such residual is 
more serious around the Galactic plane, where is highly contaminated by
the foreground emission. In order to remove the residual, it needs an aggressive 
foreground subtraction strategy that subtracts a large number of SVD modes. 
As shown in the second and third columns of Fig.~\ref{fig:mappca} and 
Fig.~\ref{fig:mappcaU}, the foreground residual is further eliminated by
subtracting $7$ or $10$ SVD modes. 

The impact of the foreground subtraction can be examined using the power spectrum. 
We estimate the angular power spectrum of each frequency slice and average 
them across the frequency band. By averaging across the frequency band, the
redshift evolution of the HI LSS is marginalized. Nevertheless, the
marginalization does not affect the examination of the foreground residual. 
The frequency-averaged angular power spectrum of the foreground cleaned maps
at different polarization leakage levels are shown in different panels of
Fig.~\ref{PCAcl}. The corresponding leakage parameters are shown in the legend 
of each panel. In each panel, the black curve represents the baseline 
HI power spectrum estimated using the input HI map. 
The blue, yellow, green, and red curves represent the recovered power spectrum 
with 3, 4, 5, and 10 SVD modes subtracted, respectively.

As shown in the first-top panel of Fig.~\ref{PCAcl}, 
without any polarization leakage, i.e. $(\epsilon_Q, \epsilon_U) = ( 0.0\%, 0.0\%)$,
the foreground can be efficiently removed by subtracting a few SVD modes. 
Meanwhile, the recovered power spectra are well below the baseline 
HI power spectrum, even with only 3 SVD modes subtracted. 
It indicates a significant signal loss, which is a common issue for 
any aggressive blind foreground subtraction.
The signal loss becomes more serious with even more SVD modes subtracted.
The same result is even more evident in the two-dimensional power spectrum. 
In Fig.~\ref{2d}, we show the $2$D power spectrum ratio of the PCA-cleaned map with respect to the input HI map
without any polarization leakage, i.e. $(\epsilon_Q, \epsilon_U) = ( 0.0\%, 0.0\%)$.
The results of PCA-$3$ and PCA-$4$ are shown in the left and right panels, respectively.  
It is clear that the PCA process incorrectly subtracts part of the HI power spectrum at the lower $k_{\parallel}$-end, including both the lower and higher $k_{\perp}$, which results in an overall signal loss
in the $C_\ell$ measurements.

When considering the polarization leakage, the foreground can not be 
efficiently removed with 3 SVD modes. As shown in the rest panels of Fig.~\ref{PCAcl}, 
the residual is more serious at large scales, i.e. $\ell \lesssim 30$. 
Similar to the results shown in the map domain, the extra foreground residual can be further 
removed by subtracting a few more SVD modes. Fig.~\ref{PCAcl} shows that
with 10 SVD mode subtraction, the foreground residual can be significantly suppressed. 
With the real HI IM survey data, the number of subtracted modes may be
much larger than 10. For example, \citet{Masui:2012zc} reports the cross-correlation 
power spectrum detection using 20 modes subtracted GBT HI IM survey maps and
\citet{2022arXiv220601579C} reports the detection of cross-correlation 
power spectrum using 30 modes subtracted MeerKAT HI IM survey maps.
The number of SVD modes needs to be subtracted depending on the complexity 
of the instrumental effect. Our simulation indicates that considering 
the instrumental effect of polarization leakage, an aggressive PCA-based foreground 
subtraction method can remove the foreground.

However, it is known that the aggressive foreground subtraction results in serious 
signal loss \citep{Switzer:2015ria}. As shown in Fig.~\ref{PCAcl}, the power spectrum 
with aggressive foreground subtraction are significantly lower than the 
baseline HI power spectrum. The signal loss can be compensated by applying 
a transfer function, which is reconstructed using simulation
\citep{Switzer:2015ria, 2022arXiv220601579C}. 
Such transfer-function-compensation method becomes the standard analysis pipeline 
applied in most of the HI IM survey data analysis. 

\subsection{Results of PCA + U-Net}
\label{R2}



\begin{figure}
\centering
\includegraphics[width=0.5\textwidth]{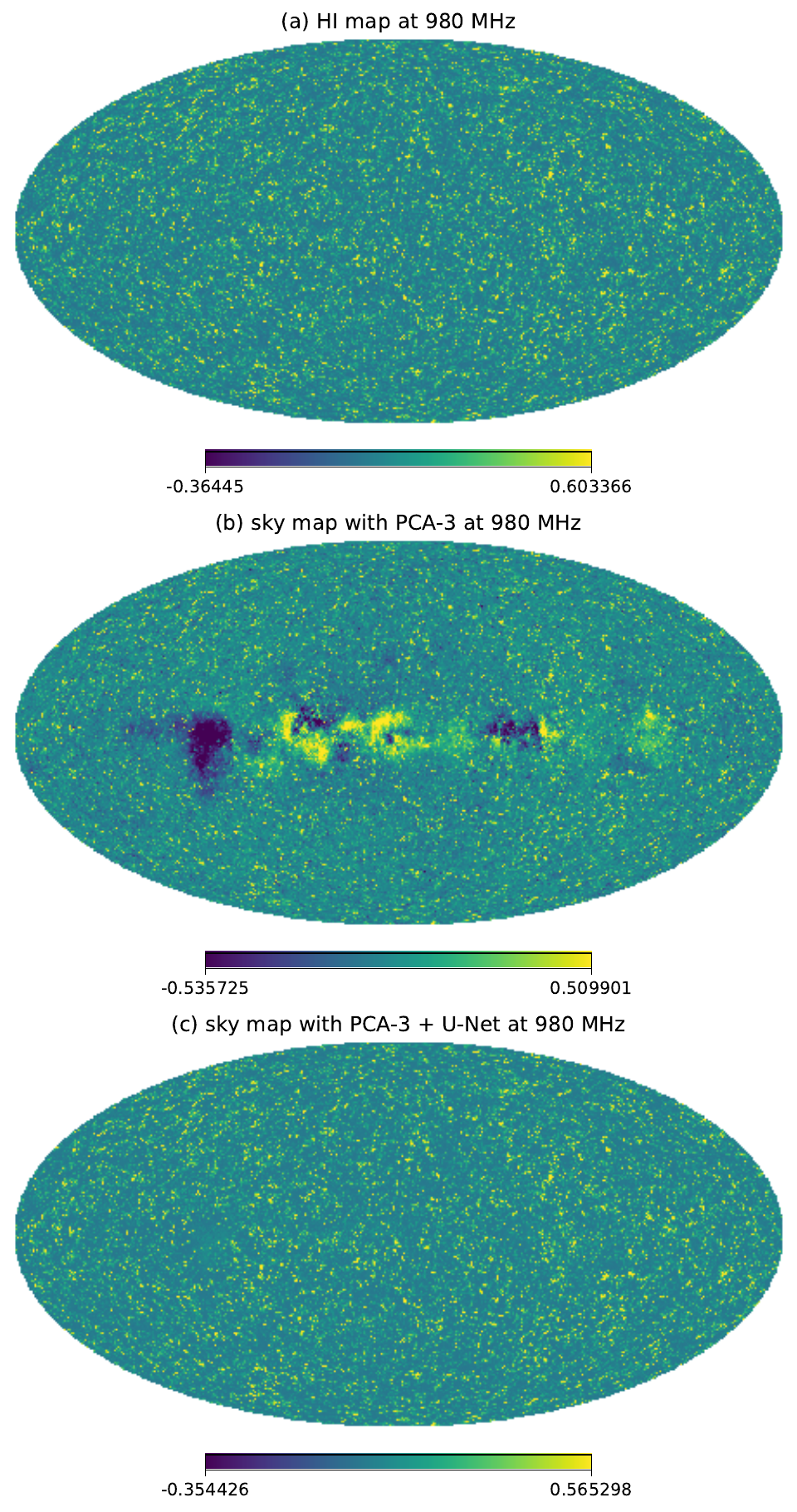}
\caption{\label{PCAHIUNet} 
The pixel-wise comparison of the foreground cleaned maps and the input HI map.
The polarization leakage parameters are assumed to be $(\epsilon_Q, \epsilon_U) = ( 0.5\%, 1.0\%)$.
Panel (a) shows the input HI map, panel (b) shows the sky map with the first 3 SVD modes subtracted, 
and panel (c) shows the sky map with additional U-Net processing after subtracting the first 3 SVD modes. 
As an example, we show only the sky maps at $980$~MHz. The unit of each pixel of the sky maps is mK.}
\end{figure}

\begin{figure}
\centering
\includegraphics[width=0.5\textwidth]{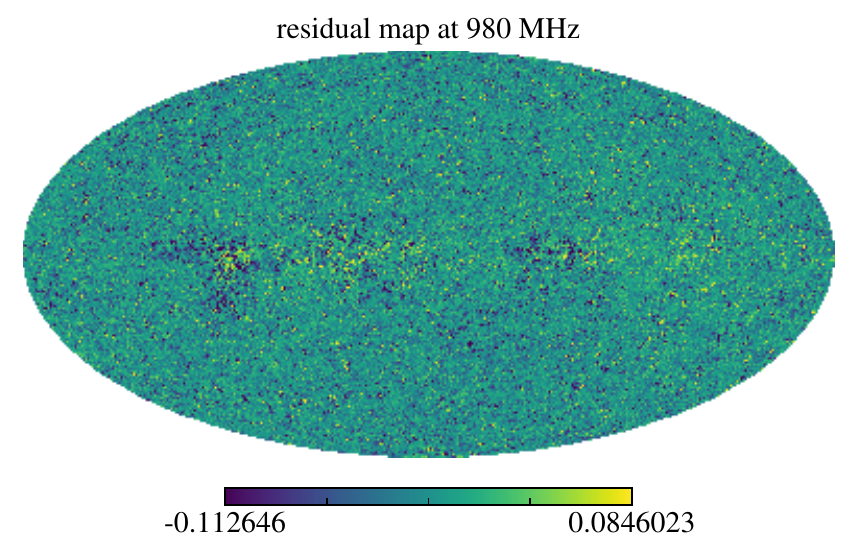}
\caption{\label{Residual} 
The differential map between the PCA-3 + U-Net cleaned map and the input HI map. 
The map uses the same polarization leakage parameter assumption as the maps shown in Fig.~\ref{PCAHIUNet},
i.e. $(\epsilon_Q, \epsilon_U) = ( 0.5\%, 1.0\%)$ and at frequency of $980$ MHz. 
The unit of each pixel of the sky maps is mK.}
\end{figure}

\begin{figure}
\centering
\includegraphics[width=0.45\textwidth]{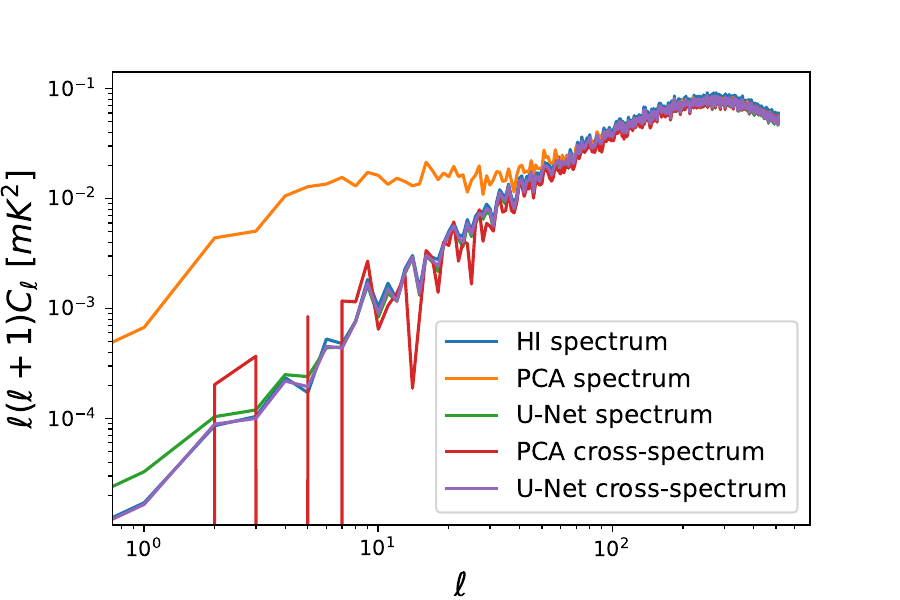}
\caption{\label{Unetcl}
Comparison of the power spectra of foreground cleaned maps with the input HI map.
The map uses the same polarization leakage parameter assumption as the maps shown in Fig.~\ref{PCAHIUNet},
i.e. $(\epsilon_Q, \epsilon_U) = ( 0.5\%, 1.0\%)$.
The power spectra are obtained using ten frequency bands averaged to eliminate errors arising from individual frequencies.
The blue curve represents the input HI signal.
The yellow and green curves represent the results of PCA-3 and PCA-3 + U-Net, and
the red and purple curves represent the corresponding cross-correlation power spectra with the input HI map.}
\end{figure}

Instead of applying aggressive foreground mode subtraction, \citep{Ni:2022kxn} proposes using the deep learning method 
to eliminate the foreground residual caused by the instrumental effect. 
According to our simulation results presented in the last section, without polarization leakage, the foreground 
contamination can be efficiently cleaned using the first 3 SVD modes. 
We assume the smooth components of the foreground can be always removed with the first 3 SVD modes
and the foreground residual due to the instrumental effect contributes to the rest SVD modes.
Before feeding the sky map to the U-Net, the PCA foreground subtraction is used as the pre-processing 
to remove the smooth components, i.e. the first 3 SVD modes. The result is shown in Fig.~\ref{PCAHIUNet}.

As an example, the reconstructed maps shown in Fig.~\ref{PCAHIUNet} use the polarization leakage 
parameter $\epsilon_Q$ and $\epsilon_U$ equal to $0.5\%$ and $1.0\%$, respectively.
The input HI sky map is shown in the top panel. 
The middle panel shows the sky maps cleaned with the first 3 SVD modes, which have significant
foreground residual around the Galactic plane.
The bottom panel shows the reconstructed sky map with the combination of PCA-3 and U-Net.
After applying the U-Net processing, the extra foreground residual structure around the Galactic plane 
is efficiently eliminated.
The amplitude range of the map is truncated between $\pm3\sigma_{\rm map}$, where $\sigma_{\rm map}$
is the rms of the map. It also shows that the reconstructed map with U-Net has similar 
rms compared to the input HI map.

In order to highlight the foreground residuals after the processing of PCA-3 + U-Net, 
Fig.~\ref{Residual} shows the differential map between the PCA-3 + U-Net cleaned map and the input HI map, 
i.e. the differential map between Fig.~\ref{PCAHIUNet}~(c) and (a).
The amplitude range of the map is also truncated between $\pm3\sigma_{\rm map}$.
Generally, the foreground residual is low and most obvious around the Galactic plane.

The corresponding angular power spectra estimated using the reconstructed maps are shown in Fig.~\ref{Unetcl}. 
As an example, we show the results with polarization leakage parameter $\epsilon_Q$ and $\epsilon_U$ equal to 
$0.5\%$ and $1.0\%$, respectively, which is the same as the maps shown in Fig.~\ref{PCAHIUNet} and Fig.~\ref{Residual}.
The angular power spectrum of the input HI map is shown with the blue curve. 
The yellow curve shows the angular power spectrum estimated using the map cleaned with only the first 3 SVD modes. 
The power spectrum estimated using the recovered HI map with the combination of PCA and U-Net is shown in green. 
It is clear that the extra power at large scales, which is induced from the polarization leakage, 
is efficiently removed with the U-Net processing. The recovered HI angular power spectrum is consistent 
with the input HI power spectrum.

To further confirm the authenticity of the reconstructed HI fluctuation, we estimate the cross-correlation 
power spectrum using the foreground cleaned map and the input HI map. 
In Fig.~\ref{Unetcl}, the cross-correlation power spectrum result using the PCA-only cleaned map is shown 
with the red curve, and the result with additional U-Net processing is shown with the purple curve.
Clearly, the large foreground residual of the PCA-only cleaned map significantly increases the variance of 
the power spectrum and results in negative values at the large-scale end. 
The large variance and negative power spectrum values also indicate the degrading of the
correlation efficiency.
However, with the additional U-Net processing, the cross-correlation angular power spectrum 
shows great agreement with the input HI power spectrum.
The good agreement between the cross- and auto-correlation power spectra indicate that the recovered HI map is 
not a fake random Gaussian field having the same power spectrum, but the truly recovered input HI map.


\begin{figure}
\centering
\begin{tikzpicture}
\scope[nodes={inner sep=0,outer sep=0}]
\node[anchor=south east] (a)
  {\includegraphics[width=0.45\textwidth]{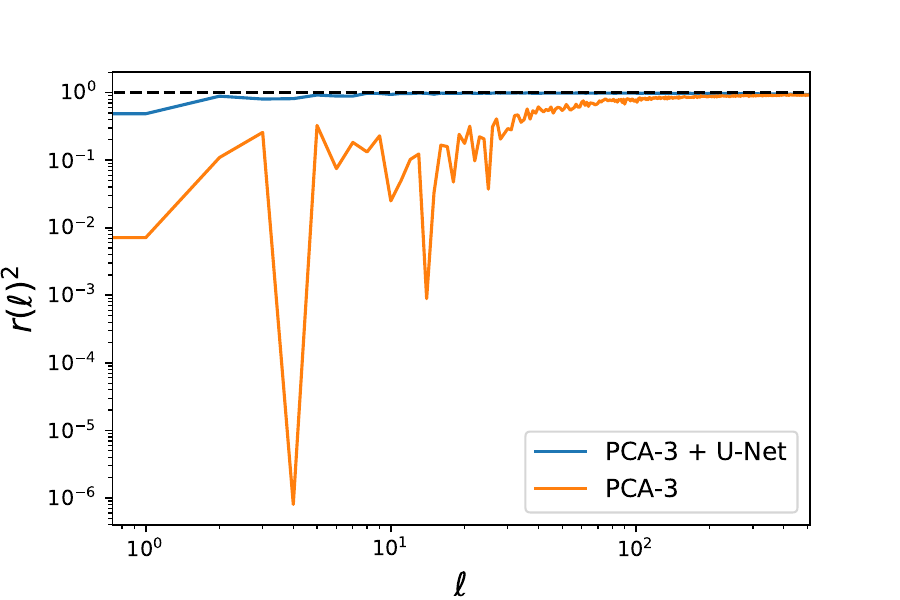}};
\node[below=0mm of a] (b)
  {\includegraphics[width=0.45\textwidth]{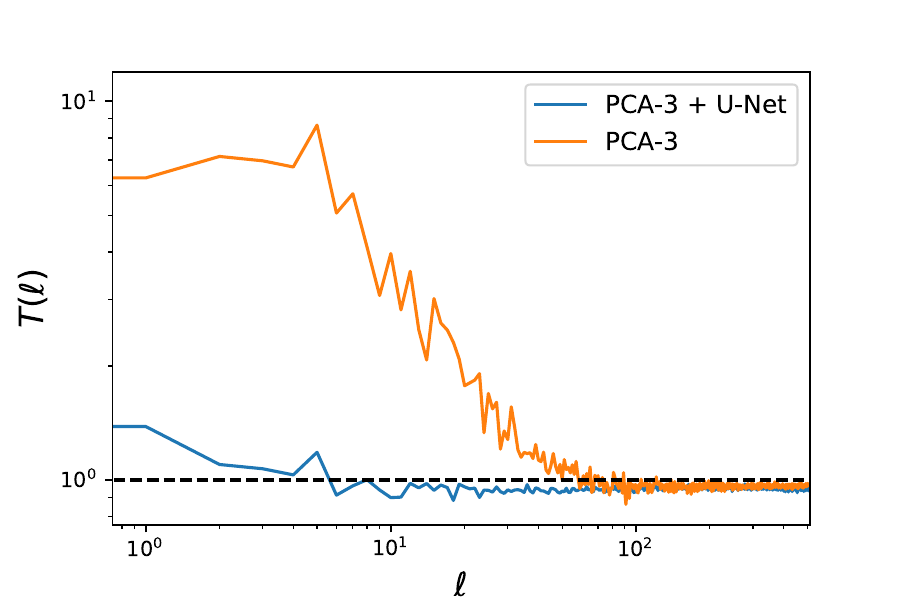}};
\endscope
\foreach \n in {a,b} {
  \node[anchor=north west] at (\n.north west) {(\n)};
}
\end{tikzpicture}
\caption{\label{spectrum_test}
Comparison of the correlation statistic of results with and without U-Net processing, assuming 3 SVD modes subtraction.
Panels (a) and (b) show the correlation statistic $r(\ell)^2$ and cleaning error $T(\ell)$, respectively. 
The results for PCA and PCA + U-Net are shown in yellow and blue curves.}
\end{figure}

In order to quantify the differences between the power spectra, we employ the correlation 
statistic and cleaning error as defined in \cite{Makinen:2020gvh}.
The correlation statistic is expressed as
\begin{equation}
r(\ell)=\frac{P_{\rm cross}(\ell)}{\sqrt{P_{\rm input}(\ell)P_{\rm rec}(\ell)}},
\label{rl}
\end{equation}
and the cleaning error is 
\begin{equation}
T(\ell)=\sqrt{\frac{P_{\rm rec}(\ell)}{P_{\rm input}(\ell)}},
\label{tl}
\end{equation}
where $P_{\rm input}(\ell)$ is the power spectrum of the input HI map, $P_{\rm rec}(\ell)$ is the power spectrum 
of the recovered HI map and $P_{\rm cross}(\ell)$ is the cross-correlation power spectrum between 
the input and recovered HI maps. The perfect recovery of the HI map results in 
$r(\ell) \sim 1$ and $T(\ell) \sim 1$. 

Using the auto- and cross-correlation power spectra shown in Fig.~\ref{Unetcl}, 
we estimate $r(\ell)$ and $T(\ell)$. The results are shown in Fig.~\ref{spectrum_test}.
In each panel of Fig.~\ref{spectrum_test}, the result of the map recovered by subtracting only 
the first 3 SVD modes is shown in yellow curve and recovered with additional U-Net 
is shown in the blue curve. Obviously, with only 3 SVD modes subtraction, the 
correlation statistic deviates from $1$ at the large-scale end. 
The deviation is mainly due to the biased auto-power spectrum and the 
large variance of the cross-correlation power spectrum at the large-scale end. 
Similarly, the cleaning error also shows a large deviation from $1$ at the large-scale end.
With the additional U-Net processing, both $r(\ell)$ and $T(\ell)$ are close to $1$.

The cleaning error $T(\ell)$ for different polarization leakage parameter settings 
are shown in Fig.~\ref{Tlall}, where Fig.~\ref{Tlall} (a) shows the results with 
different polarization leakage fraction from the Q map only and Fig.~\ref{Tlall} (b)
shows the results with additional leakage from the U map.
In both Fig.~\ref{Tlall} (a) and (b), the results with only SVD modes subtracted are shown
in the top panels and the results with extra U-Net processing are in the bottom panels.
The blue, yellow, and green curves show the result with 3, 4, and 5 SVD modes subtracted, respectively.
With only the first 3 SVD modes subtracted, 
the cleaning error shows a significant deviation from $1$ at large scales with any polarization leakage fraction.
Such deviation of cleaning error can be eliminated by subtracting more SVD modes but results in signal loss.
On the other hand, the additional U-Net processing can efficiently eliminate the foreground 
residual induced by polarization leakage. The results indicate that the 
additional U-Net processing can recover the input HI map with different levels of 
polarization leakage, as well as different levels of PCA preprocessing.

\subsection{Signal loss compensation}
\label{R3}

\begin{figure*}
\begin{tikzpicture}
\scope[nodes={inner sep=0,outer sep=5}]
\node[anchor=south east] (a)
  {\includegraphics[width=0.9\textwidth]{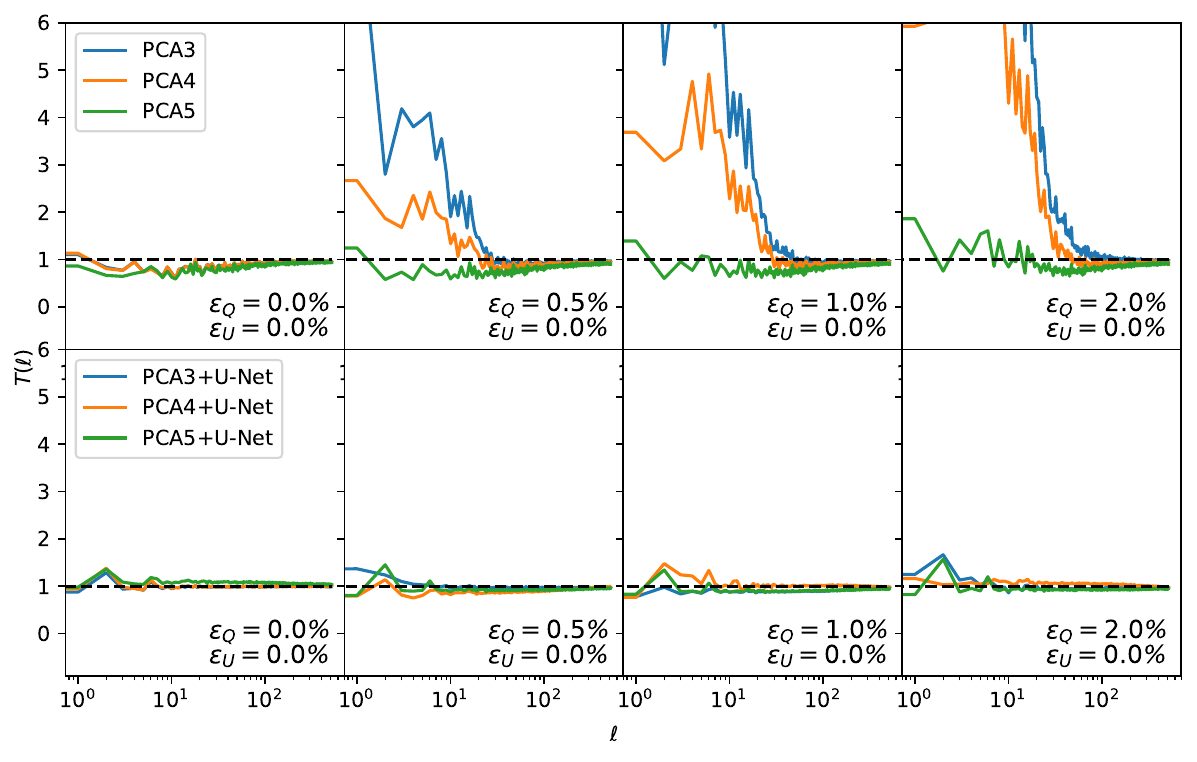}};
\node[below=0mm of a] (b)
  {\includegraphics[width=0.9\textwidth]{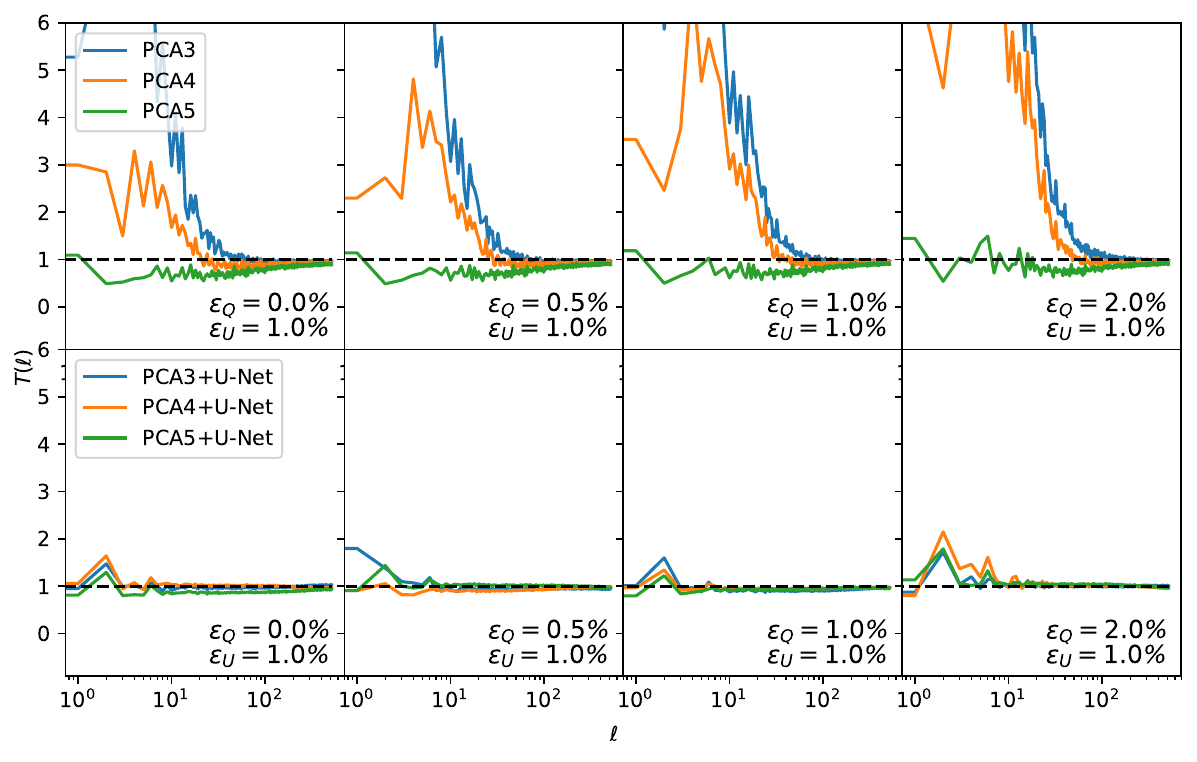}};
\endscope
\foreach \n in {a,b} {
  \node[anchor=north west] at (\n.north west) {(\n)};
}
\end{tikzpicture}
\caption{\label{Tlall}Results of the cleaning errors $T(\ell)$ obtained using U-Net for simulated sky maps of 8 different $[\epsilon_Q,\epsilon_U]$ after PCA-3, PCA-4, PCA-5 processes.
The blue, yellow, and green lines represent the input as the result of subtracting $3$, $4$, and $5$ modes from the sky maps using PCA, respectively.
The top of each of the two plots is the result of PCA and the bottom is the result of PCA+U-Net.}
\end{figure*}

In Fig.~\ref{Tlall} we also show the results with different numbers of SVD mode subtraction. 
The aggressive SVD mode subtraction can also remove the foreground leakage, but results in 
either signal loss or large variances in the power spectrum.
However, as shown in the bottom panel of Fig.~\ref{Tlall} (a) and (b), the cleaning error is 
consistent with $1$ after the additional U-Net processing. 
It means that the additional U-Net processing can also
compensate the signal killed by the aggressive PCA preprocessing.

The standard data analysis that uses PCA foreground subtraction for HI IM analysis
includes signal compensation. Typically, the signal loss is quantified by 
cleaning a set of simulated pure HI maps using the same foreground subtraction method 
as that in real data analysis. The compensation is applied
to the measurements with a transfer function, which is the power spectrum ratio 
between the cleaned and uncleaned pure HI map \citep{Switzer:2015ria,2022arXiv220601579C,Fornazier:2021ini}.
Our simulation analysis shows that such signal compensation can be replaced using
U-Net processing. Moreover, the signal compensation with U-Net processing works well 
with different numbers of SVD modes subtraction. 
It is important to note that the choice of the modes number subtracted in the PCA preprocessing
becomes adjustable with the addition of U-Net processing.
Specifically, the U-Net processing can reduce foreground residual 
when using conservative mode subtraction, while compensating for signal loss 
while using aggressive mode subtraction.

We also highlight that a considerably more complex instrumental effect may be to account
for the signal loss in the actual HI IM survey data. 
The application of actual data analysis requires a comprehensive training set that 
has as much instrumental effect as possible.

\subsection{Robustness analysis}
\label{R4}

\begin{figure}
\centering
\includegraphics[width=0.45\textwidth]{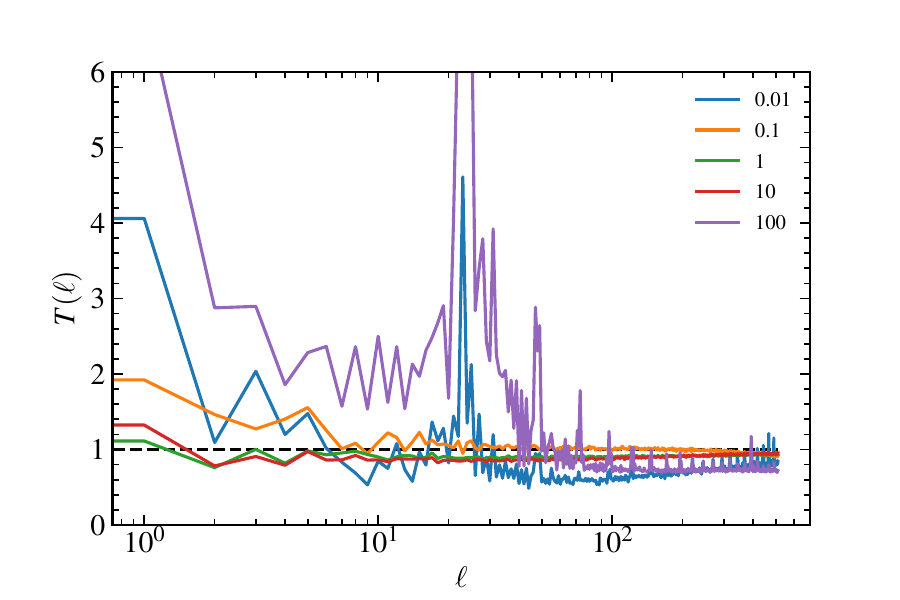}
\caption{\label{Robustnessmapall} The cleaning error $T(\ell)$ of PCA-4 + U-Net where $(\epsilon_Q, \epsilon_U) = ( 0.5\%, 1.0\%)$.
The blue, yellow, green, red, and purple lines represent the HI signal of the test sets as $0.01$, $0.1$, $1$, $10$, and $100$ times the original HI signal, respectively.}
\end{figure}

It is known that the training set for supervised deep learning techniques, such as U-Net, must 
be reasonably close to the real signal. 
However, the HI fluctuation is under-detected before the foreground contamination is removed.
To determine the robustness of the U-Net foreground subtraction, 
we aggressively assume that the true HI fluctuation could vary by a few orders 
in amplitude, i.e. multiplying the initial simulated HI maps by a factor of 
$0.01$, $0.1$, $1$, $10$, and $100$ as the test sets, respectively, and
apply the same U-Net architecture to such different test sets.
We assume the polarization leakage parameter to be $(\epsilon_Q, \epsilon_U) = ( 0.5\%, 1.0\%)$ 
and use 4 SVD modes subtracted results as the PCA preprocessing.
The cleaning error $T(\ell)$ of the different test sets are shown in Fig.~\ref{Robustnessmapall}
in different colors.
It is clear that the cleaning error increases as the HI fluctuation amplitude deviated from the 
initial HI simulation.
The results with the factor of $0.1$ and $10$ are shown with the yellow and red curves, respectively,
which indicates that the cleaning error is negligible with a minor deviation of the 
modeled HI fluctuation amplitude from the true HI signal.
However, if there is a significant deviation in the signal amplitude, 
i.e. with the factor of $0.01$ or $100$ as shown with blue or purple curve in 
Fig.~\ref{Robustnessmapall}, the U-Net foreground subtraction could result in 
large cleaning error.
Therefore, such robustness analysis infers that a successful foreground cleaning with U-Net 
processing requires the simulated HI fluctuation amplitude to be within the range of 0.1--10 times the true amplitude.
The uncertainty of the HI fluctuation amplitude, according to recent measurements, 
is less than $20\%$ \citep{2022arXiv220601579C}, which is still within the dynamical range
where U-Net foreground subtraction processing is reliable.

\section{Conclusion}
\label{sec5}

The HI IM survey is known as a promising technique for LSS studies. However, the major challenge
for HI IM survey is to remove the bright foreground contamination. 
According to the results of current HI IM survey pilot surveys, as well as the systematic analysis
with simulation, it is clear that the secret to successfully removing the bright foreground contamination
is to properly control or eliminate the instrumental effect. 
Due to the instrumental effect's complexity, it is worthwhile to investigate each one individually.
In this work, we propose a deep learning-based foreground subtraction strategy 
that can eliminate the instrumental effect injected due to systematic polarization leakage. 

We generate a set of simulated HI and foreground component maps using the open-source package {\tt CRIME}. In addition to the intensity maps, we also simulated the Stokes Q and Stokes U maps
of the Galactic synchrotron. 
In our simulation, the polarized HI fluctuation and other foreground contamination are subdominant and disregarded.
We also assume the constant polarization leakage fraction across the frequency band and
the leakage fraction from Q and U maps are parameterized as $\epsilon_Q$ and $\epsilon_U$, respectively.
The fraction of polarization leakage needs to be determined according to the observation and can be
different between different instruments. In our analysis, we generate a set of simulated sky maps
by varying the leakage fraction parameters, i.e. $\epsilon_Q = \{ 0.0\%, 0.5\%, 1.0\%, 2.0\% \}$ and
$\epsilon_U = \{0.0\%, 1.0\%\}$.

We apply the same deep learning architecture, the U-Net, as in our earlier work \citep{Ni:2022kxn} 
to eliminate polarization leakage. 
Additionally, the PCA foreground subtraction is compared and used as a preprocessing step for the 
U-Net foreground subtraction. 
Our investigation demonstrates that when polarization leakage is present, 
a conservative PCA foreground subtraction approach produces significant foreground residual, 
while an aggressive cleaning could remove the foreground residual but result in signal loss.
We also tested the differential map and cross-spectrum to ensure that the true HI signal was obtained.
On the other hand, the U-Net architecture can successfully remove foreground contamination 
at different polarization leakage levels. 

We also investigate the combination of the U-Net foreground subtraction with different
number of mode subtraction in the PCA preprocessing. 
According to our findings, the additional U-Net processing could either eliminate the 
foreground residual left over after the conservative PCA foreground subtraction or 
compensate for the signal loss caused by the aggressive PCA preprocessing.
Because of the additional U-Net processing, it becomes flexible to choose the number of modes
subtracted in the PCA preprocessing. 

Finally, we verify the robustness of the U-Net foreground subtraction strategy by varying the HI fluctuation
amplitude of the test set. The results show that 
within an amplitude range of 0.1--10 times of HI fluctuation amplitude, the U-Net subtraction strategy is reliable. 

\section*{Acknowledgements}
We thank Ming Zhang for his helpful discussions and suggestions.
We are grateful for the support of the National SKA Program of China (Grants Nos. 2022SKA0110200 and 2022SKA0110203), the National Natural Science Foundation of China (Grant Nos. 11975072 and 11835009), the Liaoning Revitalization Talents Program (Grant No. XLYC1905011), the 111 Project (Grant No. B16009), and the Top-Notch Young Talents Program of China (Grant No. W02070050).

\section*{Data Availability Statement}
The data underlying this article will be shared on reasonable request
to the corresponding author.

\bibliographystyle{mnras}
\bibliography{references.bib}

\bsp	
\label{lastpage}
\end{document}